\documentclass[a4paper,twoside,twocolumn,english]{revtex4}
\usepackage[T1]{fontenc}
\usepackage[latin1]{inputenc}
\usepackage{amsmath}
\usepackage{graphicx}
\usepackage{amssymb}

\makeatletter
\newcommand{\ket}[1]{| #1 \rangle}
\newcommand{\bra}[1]{\langle #1 |}
\newcommand{\braket}[2]{\langle #1 | #2 \rangle}

\usepackage{babel}
\makeatother
\begin{document}

\title{Fast simulation of a quantum phase transition in an ion-trap realisable unitary map}

\author{J.P. Barjaktarevic, G.J. Milburn and Ross H. McKenzie}

\address{Quantum Computer Technology Research Centre,\\
 Department of Physics, The University of Queensland,QLD 4072 Australia.}

\date{\today{}}

\begin{abstract}
We demonstrate a method of exploring the quantum critical point of
the Ising universality class using unitary maps that have recently
been demonstrated in ion trap quantum gates. We reverse the idea with
which Feynman conceived quantum computing, and ask whether a realisable
simulation corresponds to a physical system. We proceed to show that
a specific simulation (a unitary map) is physically equivalent to
a Hamiltonian that belongs to the same universality class as the transverse
Ising Hamiltonian. We present experimental signatures, and numerical
simulation for these in the six-qubit case.

\end{abstract}
\maketitle

\section{\label{sec1}Introduction}

Feynman suggested that it is possible to simulate one quantum system
with another\cite{Feynman}. However, we will turn this thesis around
by posing the question of what sort of a system some unitary map on
a quantum computer might correspond to.

In particular, we examine the ion-trap model of quantum computing,
and find that the unitary maps which have been realised on these correspond
to the time evolution of Hamiltonians which are linked closely to
the Ising model. Finally, we consider the considerable theoretical
body of work concerned with quantum phase transitions and renormalization
group theory. This will later be the key to the problem of identifying
a quantum phase transition in a unitary map.

\subsection{Simulating Quantum Systems}

Feynman's first conception of quantum computing\cite{Feynman} held
the simulation of quantum systems as a key goal. Entanglement has
been described as the quintessential feature of quantum mechanics\cite{Schrodinger}.
In general, the arbitrary time evolution of a system is considered
an \emph{NP-hard\cite{NP}} problem, as memory and processing resources
increase exponentially in the size of the problem, $n$, on a classical
computer. It is only for extraordinarily simple systems, or ones for
which there are strong symmetries, that such calculations are tractable.

Feynman suggested that the problem could be reduced to one in polynomial
time on a computer based on quantum principles. These include the
ability of a quantum system to perform unitary operations on a set
of quantum bits (\emph{qubits}), and to exist in entangled states.
Feynman showed that, in principle, it was possible to perform, in
polynomial time, algorithms which were only possible in non-polynomial
time on a classical computer.

Since the original formulation of the problem, the application of
quantum computing to classical problems has become more common. Several
algorithms have been suggested, including the Deutch-Jozsa algorithm\cite{DJ},
Shor's factorization algorithm\cite{Shor}, and Grover's searching
algorithm\cite{Grover}. However, all of these systems are widely
considered far removed from current experimental abilities.

Recently, Lloyd\cite{Lloyd} revisited Feynman's original problem,
and showed that it was possible to implement the time evolution of
an arbitrary spin Hamiltonian to a particular precision, $\varepsilon$,
in polynomial time. The procedure essentially involves the decomposition
of a Hamiltonian into realizable (local) unitary operations, and the
time-wise stepping through a Hamiltonian to some arbitrary accuracy.

It will be our desire to avoid such an abstracted simulation of a
quantum system, and rather consider the possibility of finding a quantum
phase transition in a quantum algorithm naturally realizable with
current quantum computing experimental hardware. In this way, we will
essentially reverse the Feynman thesis, and conclude that quantum
algorithms (or unitary maps) will correspond to the observables of
some physical system.

\subsection{Ion Trap Quantum Computers and the Ising Model}

DiVincenzo\cite{Universal} and Barenco \emph{et al.}\cite{Universal2}
has shown that single-site rotations and two-site controlled NOTs
are universal for quantum computation. Further, the Sorensen-Molmer\cite{smgate},
phase gate\cite{Wineland}, and indeed almost any two-site entangling
gate\cite{Nielsen} are universal. Hence, they will be able to affect
any unitary transformation.

Cirac and Zoller's paper\cite{Ion} on cold ion-trap quantum computers
introduces the use of a spatially confined ion spin as a qubit, and
the excitation of vibrational modes as a means of coupling qubits.
Further, it has been shown that high fidelity state-preparation\cite{hifiprep}
and readout\cite{hifireadout} are feasible.

Milburn has suggested a robust phase space scheme to use ion traps
to simulate nonlinear interactions in spin systems\cite{milburn}.
A significant advantage of this scheme is that it does not require
the cooling of vibrational states. The method involves the application
of Raman pulses faster than the vibrational heating time, effectively
decoupling the effect of vibrational modes. In particular, Milburn
shows that the evolution of a Hamiltonian of the form

\begin{equation}
H_{int}=\hbar\chi\sigma_{z}^{(1)}\sigma_{z}^{(2)}\label{eq:}\end{equation}
may be achieved by a pulse sequence

\begin{eqnarray}
U_{int}=e^{-iH_{int}} & = & e^{i\kappa_{x}\hat{X}\sigma_{z}^{(1)}}e^{i\kappa_{p}\hat{P}\sigma_{z}^{(2)}}\label{eq:}\\
 &  & e^{-i\kappa_{x}\hat{X}\sigma_{z}^{(1)}}e^{-i\kappa_{p}\hat{P}\sigma_{z}^{(2)}}\nonumber \end{eqnarray}
where $\hat{X}=\frac{a+a^{\dagger}}{\sqrt{2}}$ and $\hat{P}=\frac{a-a^{\dagger}}{i\sqrt{2}}$,
and expressions for $\kappa_{x}$ and $\kappa_{p}$ given in Ref.
\cite{warm_milburn}.

Further, Wineland's research group have recently demonstrated\cite{Wineland}
considerable success in achieving few-qubit interactions with this
scheme. In particular, they present a two qubit phase gate, which
has the form

\begin{eqnarray*}
\left|\downarrow\downarrow\right\rangle \rightarrow\left|\downarrow\downarrow\right\rangle  & , & \left|\uparrow\uparrow\right\rangle \rightarrow\left|\uparrow\uparrow\right\rangle ,\\
\left|\downarrow\uparrow\right\rangle \rightarrow e^{i\phi}\left|\downarrow\uparrow\right\rangle  & . & \left|\uparrow\downarrow\right\rangle \rightarrow e^{i\phi}\left|\uparrow\downarrow\right\rangle \end{eqnarray*}
which can be recast as $\left|\Psi\right\rangle \rightarrow e^{-i\chi\sigma_{x}^{(1)}\sigma_{x}^{(2)}}\left|\Psi\right\rangle $.
Apart from an uninteresting global additive phase, this may be considered
to model the time evolution of a Hamiltonian of the form $\sigma_{x}^{(n)}\sigma_{x}^{(n+1)}$.

Further, it is well known that single rotations in any basis, which
correspond to the evolution of a spin operator, are easily implementable
on such an architecture\cite{Ion}. They result in unitary transformations
of the form

\begin{equation}
U_{single}=e^{-i\hbar\theta\sigma_{x}^{(1)}}\label{eq:}\end{equation}
which can be implemented trivially through a single Raman pulse.

Following Feynman's original intentions for quantum computing, one
may consider the mapping of Hamiltonian with such terms onto an ion-trap
quantum computer. Turning this problem around, we will consider the
properties of a unitary map composed of terms which can be experimentally
implemented, and investigate their relationship with the transverse
Ising spin chain.

\subsection{Quantum Phase Transitions and Universality Classes}

The quantum phase transition in the one dimensional transverse Ising
model\cite{Sachdev} is very well understood. The Hamiltonian is given
by:

\begin{equation}
H_{Ising}=\sum_{n=1}^{N}\mu B\sigma_{x}^{(n)}+J\sigma_{z}^{(n)}\sigma_{z}^{(n)}\label{ising}\end{equation}

It is known that for an external field with interaction strength \textbf{$\mu B$}
and local exchange interaction term with strength $J$, that a phase
transition occurs for $\mu B=\pm J$. One can intuitively consider
the phase transition as a result of the incongruent symmetries between
the two phases, which is reflected in the difference in behaviour
of the two terms in the Hamiltonian under the transformation $\sigma_{n}\rightarrow-\sigma_{n}$.
In the regime $J>\mu B$, the system is in a ferromagnetic phase,
with $\left\langle \sigma_{x}^{(n)}\right\rangle \neq0$, and the
system displays long range order. On the other hand, for $J<\mu B$,
the system is paramagnetic, with$\left\langle \sigma_{x}^{(n)}\right\rangle =0$,
and there is no broken symmetry.

Using arguments from renormalization group theory, we may place a
great number of related problems into the same universality class\cite{Cardy},
and we may expect to see a similar phase transition occur in a number
of related systems.

\section{The Model}

In the following, we will put together the components introduced in
Section \ref{sec1} in a intuitive way. We consider the composition
of the two unitary maps, similar to those demonstrated in Ref \cite{Wineland},
which corresponds to the composition of the time evolution of two
Hamiltonians. In form, it will look similar to the one-dimensional
transverse Ising chain Hamiltonian. We will then apply the Jordan-Wigner
transformation to this model to express the Hamiltonians in terms
of non-interacting fermions. We are then able to perform a composition
of operators in an $SU(2)$ representation to yield a single Hamiltonian.
We will find that this model is highly non-local. However, using renormalization
group theory concepts, it can be shown that the Hamiltonian belongs
in the same universality class as the transverse Ising chain. Hence,
we conclude that our separated model has the same quantum phase transition
as the transverse Ising chain, even though we have implemented the
map in a much simpler way.

\subsection{Model Unitary Transformation and Experimental Realization}

It is natural to decompose the Ising Hamiltonian, $H_{Ising}$ into
two distinct parts:

\begin{equation}
H_{\chi}=\chi\sum_{n=1}^{N}\sigma_{z}^{(n)}\sigma_{z}^{(n+1)}\label{eq:}\end{equation}

\begin{equation}
H_{\theta}=\theta\sum_{n=1}^{N}\sigma_{x}^{(n)}\label{eq:}\end{equation}

These parts are of even and odd symmetry under $\vec{\sigma_{n}}\rightarrow-\vec{\sigma_{n}}$,
respectively. Unitary maps of the form $\left|\Psi\right\rangle \rightarrow e^{iH_{\chi,\theta}}\left|\Psi\right\rangle $
have been realised experimentally. It is impossible to perform them
both at the same time with only single qubit rotations and two qubit
gates, because they do not commute - the evolution of the combined
Hamiltonian is not the composition of the evolutions of both Hamiltonians.
Note however that terms $\sigma_{z}^{(n)}\sigma_{z}^{(n+1)}$ and
$\sigma_{z}^{(m)}\sigma_{z}^{(m+1)}$ do commute, and so

\begin{equation}
e^{iH_{\chi}}=\prod_{n=1}^{N}e^{i\chi\sigma_{z}^{(n)}\sigma_{z}^{(n+1)}}\label{eq:}\end{equation}
is realisable in principle with current technology.

The combined Hamiltonian may be approximated by Lloyd's\cite{Lloyd}
methods, which involves applying terms such as $\frac{1}{m}H_{\chi}$
and $\frac{1}{m}H_{\theta}$ repeatedly, $m$ times. However this
requires a large overhead - instead we will consider the unitary map

\begin{equation}
U(\chi,\theta)=e^{-iH_{\chi}}e^{-iH_{\theta}}=e^{-i\bar{H}}\neq e^{-i(H_{\chi}+H_{\theta})}\label{u_def}\end{equation}

This map has been proposed by Milburn \emph{et al.} \cite{warm_milburn}
as an easier unitary map to to simulate than the map which corresponds
to the time evolution of transverse Ising chain Hamiltonian. We are
interested in whether this mapping will have the same quantum phase
transition behaviour as the transverse Ising chain.

\subsection{Jordan-Wigner Transformation}

We will follow Jordan and Wigner\cite{JW} in using the following
definitions to introduce a new set of operators, $a_{n}$, where

\begin{eqnarray}
\sigma_{x}^{(n)} & = & 1-2a_{n}a_{n}^{\dagger}\label{eq:}\\
\sigma_{y}^{(n)} & = & -i(a_{n}-a_{n}^{\dagger})\label{eq:}\\
\sigma_{z}^{(n)} & = & a_{n}^{\dagger}+a_{n}\label{eq:}\end{eqnarray}
where $\sigma_{x}^{(n)}$, $\sigma_{y}^{(n)}$ and $\sigma_{z}^{(n)}$
take the form of the Pauli spin matrices in the $\left|0\right\rangle $,$a_{n}^{\dagger}\left|0\right\rangle $
basis. From these definitions, the operators $a_{n}$ and $a_{n}^{\dagger}$
can be shown to obey the following relations:

\begin{eqnarray*}
\{ a_{n}^{\dagger},a_{n}\}=1, & a_{n}^{2}=0, & a_{n}^{\dagger^{2}}=0,\\
{}[a_{m}^{\dagger},a_{n}]=0, & [a_{m}^{\dagger},a_{n}^{\dagger}]=0, & [a_{m},a_{n}]=0,m\ne n\end{eqnarray*}

With these definitions, our unitary map becomes

\begin{eqnarray}
U(\chi,\theta) & = & e^{-i\chi\sum_{n=1}^{N}a_{n}^{\dagger}a_{n+1}^{\dagger}+a_{n}a_{n+1}+a_{n}a_{n+1}^{\dagger}+a_{n}^{\dagger}a_{n+1}}\label{u_def_c}\\
 &  & e^{-i\theta\sum_{n=1}^{N}1-2a_{n}a_{n}^{\dagger}}\nonumber \end{eqnarray}

We then introduce the following operators

\begin{eqnarray}
c_{n} & = & e^{i\pi\sum_{j=1}^{n-1}a_{j}^{\dagger}a_{j}}a_{n}\label{eq:}\\
c_{n}^{\dagger} & = & a_{n}^{\dagger}e^{-i\pi\sum_{j=1}^{n-1}a_{j}^{\dagger}a_{j}}\label{eq:}\end{eqnarray}

It can be shown that they obey fermionic anti-commutation relations.

We may understand these as an expression of domain wall creation and
destruction. We can re-express $U(\chi,\theta)$ with this new set
of operators as

\begin{eqnarray}
U(\chi,\theta) & = & e^{-i\chi\sum_{n=1}^{N}c_{n}^{\dagger}c_{n+1}^{\dagger}-c_{n}c_{n+1}-c_{n}c_{n+1}^{\dagger}+c_{n}^{\dagger}c_{n+1}}\label{eq:}\\
 &  & e^{-i\theta\sum_{n=1}^{N}c_{n}^{\dagger}c_{n}-c_{n}c_{n}^{\dagger}}\nonumber \end{eqnarray}

Finally, we will define the Fourier transformed versions of the fermion
operators as

\begin{eqnarray}
c_{n} & = & \frac{1}{\sqrt{N}}\sum_{k}C_{k}e^{ink}\label{eq:}\\
c_{n}^{\dagger} & = & \frac{1}{\sqrt{N}}\sum_{k}C_{k}^{\dagger}e^{-ink}\label{eq:}\end{eqnarray}

However, it is important to take note of the boundary terms. Strictly,
in order to have Eqs. (\ref{u_def}) and (\ref{u_def_c}) identical,
we must make the identification\cite{cyclic}

\[
c_{N+1}=c_{1}(e^{i\sum_{j=1}^{N}c_{j}^{\dagger}c_{j}}+1)\]

It may be argued that in the thermodynamic limit, this term will be
irrelevant, and we may make the identification $c_{N+1}=c_{1}$.

Due to cyclic boundary conditions, we will require $k$ to take the
discrete values

\[
k=\frac{2\pi m}{L},m=-\frac{L}{2},...,-1,0,1,\frac{L-2}{2}\]

These operators satisfy fermion anti-commutation relations:

\begin{eqnarray*}
\{ C_{k},C_{l}^{\dagger}\} & = & \delta_{kl}\\
\{ C_{k},C_{l}\} & = & \{ C_{k}^{\dagger},C_{l}^{\dagger}\}=0\end{eqnarray*}

Using the definitions of $c_{n}$ and $c_{n}^{\dagger}$, and the
thermodynamic limit we can re-write $U(\chi,\theta)$ as

\begin{eqnarray}
U(\chi,\theta) & = & e^{-i\chi\sum_{k}2\cos kC_{k}^{\dagger}C_{k}-i\sin k(C_{k}^{\dagger}C_{-k}^{\dagger}+C_{k}C_{-k})}\label{eq:}\\
 &  & e^{-i\theta\sum_{k}(2C_{k}^{\dagger}C_{k}-1)}\nonumber \end{eqnarray}
where we require the thermodynamic limit so that the property $2\sum_{k}C_{k}^{\dagger}C_{k}=\sum_{k}(C_{k}^{\dagger}C_{k}+C_{-k}^{\dagger}C_{-k})$
holds.

To simplify matters, let us further define

\begin{eqnarray}
\hat{A}_{k} & = & \chi(2\cos kC_{k}^{\dagger}C_{k}-i\sin k(C_{k}^{\dagger}C_{-k}^{\dagger}+C_{k}C_{-k}))\label{eq:}\\
\hat{B}_{k} & = & \theta(2C_{k}^{\dagger}C_{k}-1)\label{eq:}\end{eqnarray}
such that we may write

\begin{equation}
U(\chi,\theta)=e^{-i\sum_{k}\hat{A}_{k}}e^{-i\sum_{k}\hat{B}_{k}}=\Pi_{k}U_{k}(\chi,\theta)\label{eq:}\end{equation}
where $U_{k}(\chi,\theta)\equiv e^{-i\hat{A}_{k}}e^{-i\hat{B}_{k}}$.

We have now completely decoupled the problem, and may express the
operators $\hat{A_{k}}$ and $\hat{B_{k}}$ in the basis $\left|0\right\rangle ,C_{k}^{\dagger}\left|0\right\rangle ,C_{-k}^{\dagger}\left|0\right\rangle ,C_{k}^{\dagger}C_{-k}^{\dagger}\left|0\right\rangle $.
It is be possible to find eigenstates of $U(\chi,\theta)$ in closed
form in this basis.

\subsection{Combining }

However, it would be nice to be able to express $U(\chi,\theta)$
as a single exponential. While $\hat{A_{k}}$ and $\hat{B_{k}}$ do
not commute, it turns out that there is a faithful representation
in $SU(2)$, if we make the following definitions : 

\begin{eqnarray*}
\nu_{1}^{(k)} & = & -i(C_{k}^{\dagger}C_{-k}^{\dagger}+C_{k}C_{-k})\\
\nu_{2}^{(k)} & = & (-C_{k}^{\dagger}C_{-k}^{\dagger}+C_{k}C_{-k})\\
\nu_{3}^{(k)} & = & C_{k}^{\dagger}C_{k}+C_{-k}^{\dagger}C_{-k}-I\end{eqnarray*}

Hence, we can express $\hat{A}_{k}=\chi(\cos k+\vec{\alpha}.\vec{\nu_{k}})$
and $\hat{B}_{k}=\theta\vec{\beta}_{k}.\vec{\nu}_{k}$, where $\vec{\alpha}_{k}=\chi(\sin k,0,\cos k)$
and $\vec{\beta}_{k}=\theta(0,0,1)$. We have that $[\nu_{l}^{(k)},\nu_{m}^{(k')}]=-2i\epsilon_{l,m,n}\nu_{n}^{(k)}\delta_{k}^{k'}$
where $\epsilon_{l,m,n}$ is the Levi-Civita symbol, so that $\{\nu_{1}^{(k)},\nu_{2}^{(k)},\nu_{3}^{(k)}\}$
have the same properties as the $SU(2)$ matrices $\{\sigma_{1},\sigma_{2},\sigma_{3}\}$.
Relating the fermionic operators to $SU(2)$ in this way was inspired
by a similar approach in the theory of superconductors\cite{Schreiffer}.

$SU(2)$ is closed under composition with a well understood composition
relation, which we can now apply to our system\cite{closed2}

\begin{equation}
U_{k}(\chi,\theta)=e^{-i\chi\vec{\alpha}_{k}.\vec{\nu_{k}}}e^{-i\theta\vec{\beta}_{k}.\vec{\nu_{k}}}=e^{-i\cos k}e^{-i\kappa_{k}\vec{\gamma}_{k}(\chi,\theta).\vec{\nu_{k}}}\label{u_final}\end{equation}
where 

\begin{eqnarray}
\vec{\gamma}_{k}(\chi,\theta) & = & (\sin k\cos\theta\sin\chi,-\sin k\sin\theta\sin\chi,\label{gamma_def}\\
 &  & (\sin\theta\cos\chi+\cos k\cos\theta\sin\chi))\nonumber \\
\kappa_{k} & = & {\frac{\cos^{-1}\eta_{k}}{\sqrt{1-\eta_{k}^{2}}}}\label{kappa_def}\\
\eta_{k} & = & \cos\theta\cos\chi-\cos k\sin\theta\sin\chi\label{eta_def}\\
 & = & \cos^{2}\frac{k}{2}\cos(\theta+\chi)+\sin^{2}\frac{k}{2}\cos(\theta-\chi)\nonumber \end{eqnarray}

This composition has the simple physical interpretation of two rotations
being composed, and the result can be derived using quaternion composition\cite{closed}.
However, when using quaternions, special care has to be given to the
double cover of $SO(3)$ under $SU(2)$. The second equality of Eq.
(\ref{u_final}) defines an effective Hamiltonian, $\bar{H}_{k}$,
and we stress that $\bar{H}_{k}\neq\hat{A}_{k}+\hat{B}_{k}$ because
$\hat{A}_{k}$and $\hat{B}_{k}$ do not commute.

Hence, we have the final form of the decoupled, and combined transformation

\begin{equation}
U(\chi,\theta)=\Pi_{k}U_{k}(\chi,\theta)=e^{-i\sum_{k}\kappa_{k}\vec{\gamma}_{k}(\chi,\theta).\vec{\nu_{k}}}\label{decoupled}\end{equation}

Hence, we have found that the effective Hamiltonian , $\bar{H}$,
defined in Eq (\ref{u_def}) is given by $\bar{H}=\sum_{k}\kappa_{k}\vec{\gamma}_{k}(\chi,\theta).\vec{\nu_{k}}$

We now check the limit $\chi\rightarrow0$, which implies $\kappa\rightarrow{\frac{\theta}{\sin\theta}}$,
and $\vec{\gamma}_{k}\rightarrow\{0,0,\cos k\sin\theta\}$. Hence
$U(\chi,\theta)=e^{-i\sum_{k}{\frac{\theta}{\sin\theta}}\sin\theta\cos k\nu_{3}}=e^{-i\theta\sum_{k}{\frac{-i}{2}}\sigma_{x}\sigma_{y}-\sigma_{y}\sigma_{x}}=e^{-i\theta\sum_{k}{\frac{-i}{2}}2i\sigma_{x}\sigma_{y}}=e^{-i\theta\sum_{k}\sigma_{z}}$.
On the other hand, in the limit $\theta\rightarrow0$, $\kappa={\frac{\chi}{\sin\chi}}$,
and $\vec{\gamma}_{k}=\{\sin\chi\sin k,0,\sin\chi\cos k\}$. Hence
$U(\chi,\theta)=e^{-i\chi\sum_{k}\sin k\nu_{1}+\cos_{k}\nu_{3}}$.
Thus, we retrieve the expected behaviour in the limit as we turn off
either the exchange or external field terms.

Having expressed $U(\chi,\theta)$ in this form, it is now possible
to show that it directly corresponds to some physical Hamiltonian.
We may perform a Bogoliubov transformation by defining some fermion
creation operator

\begin{equation}
\gamma_{k}\gamma_{k}^{\dagger}=\vec{\gamma}_{k}(\chi,\theta).\vec{\nu_{k}}\label{eq:}\end{equation}
with associated energy, $\epsilon_{k}=\kappa_{k}$. Hence, we may
consider our ground state as a vacuum state $|0>$, and excitations
as $\gamma_{k}^{\dagger}\left|0\right\rangle $. It is important to
note here that the excitations of lowest energy will occur at an extremum
of $\epsilon_{k}$. We can show that this occurs at $k=0,\pi$ by
noting that

\begin{equation}
\frac{\partial\epsilon_{k}}{\partial k}=\frac{\partial\kappa_{k}}{\partial k}=\frac{\partial\kappa_{k}}{\partial\eta_{k}}\frac{\partial\eta_{k}}{\partial k}\label{eq:}\end{equation}
from which it follows that

\begin{equation}
\left.\frac{\partial\eta_{k}}{\partial k}\right|_{k=0,\pi}=\left.\sin k\sin\theta\sin\chi\right|_{k=0,\pi}=0\label{eq:}\end{equation}

Hence, the elementary excitations will be for $k=0$ or $k=\pi$,
whichever corresponds to a lower energy.

\subsection{Closed-form Hamiltonian}

We can now work backwards from our expression for $U(\chi,\theta)$
to a single combined Hamiltonian. The results here will only be valid
in the thermodynamic limit, which we have assumed in the previous
section. Before doing so, we should present a list of identities which
will prove to be useful.

\begin{eqnarray*}
\sum_{k}e^{iak}\nu_{1}^{(k)} & = & i\sum_{n}c_{n}^{\dagger}c_{n+a}^{\dagger}-c_{n}c_{n+a}\\
\sum_{k}e^{iak}\nu_{2}^{(k)} & = & -\sum_{n}c_{n}^{\dagger}c_{n+a}^{\dagger}+c_{n}c_{n+a}\\
\sum_{k}e^{iak}\nu_{3}^{(k)} & = & \sum_{n}2c_{n}^{\dagger}c_{n}-I\\
\sum_{k}\nu_{1}^{(k)}=0 & , & \sum_{k}\cos(ak)\nu_{1}^{(k)}=0\\
\sum_{k}\sin(ak)\nu_{1}^{(k)} & = & \sum_{n}c_{n}^{\dagger}c_{n+a}^{\dagger}-c_{n}c_{n+a}\\
\sum_{k}\nu_{2}^{(k)}=0 & , & \sum_{k}\cos(ak)\nu_{2}^{(k)}=0\\
\sum_{k}\sin(ak)\nu_{2}^{(k)} & = & i\sum_{n}(c_{n}^{\dagger}c_{n+a}^{\dagger}+c_{n}c_{n+a})\end{eqnarray*}

Now, recall that 

\begin{equation}
U(\chi,\theta)=e^{-i\sum_{k}\kappa_{k}\vec{\gamma}_{k}(\chi,\theta).\vec{\nu}_{k}}\label{eq:}\end{equation}

However, $\kappa_{k}$ is an even function of $k$, and so can be
expanded in terms of a Fourier series involving $\cos k$. Let us
write

\begin{equation}
\kappa_{k}=\sum_{l=0}^{\infty}a_{l}\cos(lk)\label{eq:}\end{equation}

\begin{equation}
U(\chi,\theta)=e^{-i\sum_{k,l}a_{l}\cos(lk)\vec{\gamma}_{k}(\chi,\theta).\vec{\nu}_{k}}\label{a_series}\end{equation}

We will now substitute our expression for $\vec{\gamma}_{k}$ and
expand.

\begin{eqnarray}
U(\chi,\theta) & = & e^{-i\sum_{k,l}a_{l}\cos lk\vec{\gamma_{k}}(\chi,\theta).\vec{\nu}_{k}}\label{eq:}\\
 & = & e^{-i(\Lambda_{1}+\Lambda_{2}+\Lambda_{3})}\label{eq:}\\
 & = & e^{-i\bar{H}}\label{eq:}\end{eqnarray}

where

\begin{eqnarray}
\Lambda_{1} & = & \cos\theta\sin\chi\sum_{k,l}a_{l}\sin k\cos(lk)\nu_{1}^{(k)}\label{eq:}\\
\Lambda_{2} & = & -\sin\theta\sin\chi\sum_{k,l}a_{l}\sin k\cos(lk)\nu_{2}^{(k)}\label{eq:}\\
\Lambda_{3} & = & \sin\theta\cos\chi\sum_{k,l}a_{l}\cos(lk)\nu_{3}^{(k)}\label{eq:}\\
 &  & +\cos\theta\sin\chi\sum_{k,l}a_{l}\cos k\cos(lk)\nu_{3}^{(k)}\nonumber \end{eqnarray}

and the sum over $l$ ranges $1,2,3...$

We may rewrite this as

\begin{eqnarray}
\Lambda_{1} & = & \cos\theta\sin\chi\sum_{k,l}a_{l}[\sin(l+1)k-\sin(l-1)k]\nu_{1}^{(k)}\label{eq:}\\
 & = & \cos\theta\sin\chi[a_{0}(c_{n}^{\dagger}c_{n+1}^{\dagger}-c_{n}c_{n+1})\label{eq:}\\
 &  & +\sum_{n,l}\frac{(a_{l+1}-a_{l-1})}{2}(c_{n}^{\dagger}c_{n+l}^{\dagger}-c_{n}c_{n+l})]\nonumber \\
\Lambda_{2} & = & -\sin\theta\sin\chi\sum_{k,l}a_{l}[\sin(l+1)k-\sin(l-1)k]\nu_{2}^{(k)}\label{eq:}\\
 & = & -i\sin\theta\sin\chi\sum_{n,l}[a_{o}(c_{n}^{\dagger}c_{n+1}^{\dagger}+c_{n}c_{n+1})\label{eq:}\\
 &  & +\sum_{l}\frac{(a_{l+1}-a_{l-1})}{2}(c_{n}^{\dagger}c_{n+l}^{\dagger}+c_{n}c_{n+l})]\nonumber \\
\Lambda_{3} & = & \sin\theta\cos\chi\sum_{k,l}a_{l}\cos(lk)\nu_{3}^{(k)}\label{eq:}\\
 &  & +\cos\theta\sin\chi\sum_{k,l}a_{l}[\cos(l+1)k+\cos(l-1)k]\nu_{3}^{(k)}\nonumber \\
 & = & \sin\theta\cos\chi\sum_{l}[a_{l}(c_{n}^{\dagger}c_{n+l}-c_{n}c_{n+l}^{\dagger})]+\label{eq:}\\
 &  & +\cos\theta\sin\chi\sum_{n,l}[a_{0}(c_{n}^{\dagger}c_{n+1}-c_{n}c_{n+1}^{\dagger})\nonumber \\
 &  & +\frac{(a_{l+1}-a_{l-1}))}{2}(c_{n}^{\dagger}c_{n+l}-c_{n}c_{n+l}^{\dagger})]\nonumber \end{eqnarray}

Therefore, the quantum spin chain Hamiltonian $\bar{H}$ which represents
a physical system corresponding to the separated unitary map (\ref{u_def})
is highly non-local. Terms such as $c_{n}c_{n+a}^{\dagger}=a_{n}e^{-i\pi\sum_{j=n}^{n+a-1}a_{j}a_{j}^{\dagger}}a_{n+a}^{\dagger}$
for $a>1$ will not only involve $a_{n},a_{n+a}$, but also $c_{m}$
and $c_{m}^{\dagger}$ $\forall n<m<n+a$.

If we define

\begin{eqnarray*}
\sigma_{+}^{(n)} & = & \frac{\sigma_{z}^{(n)}+i\sigma_{y}^{(n)}}{2}\\
\sigma_{-}^{(n)} & = & \frac{\sigma_{z}^{(n)}-i\sigma_{y}^{(n)}}{2}\end{eqnarray*}
we can write $c_{n}c_{n+a}^{\dagger}$ as $\sigma_{+}^{(n)}e^{-i\pi\sum_{j=n}^{n+a-1}\frac{1-\sigma_{x}^{(j)}}{2}}\sigma_{-}^{(n+a)}$,
explicitly showing the dependence on non-neighbouring spins.

\subsection{Range of the Interactions}

\begin{figure}
\includegraphics[%
  width=3.04414in]{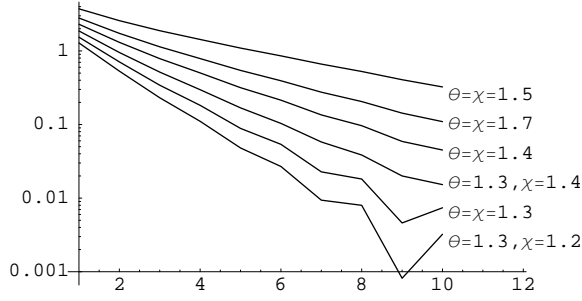}

\caption{\label{a_coeff}The behaviour of Fourier coefficients $a_{n}$, as
defined in Eq. \ref{a_series}, for $n=1,2...10$ for a variety of
$\theta$ and $\chi$. Note that the larger $\left|\chi-\frac{\pi}{2}\right|$
and $\left|\theta-\frac{\pi}{2}\right|$, the larger the decay. The
exponential decay in these coefficients implies that the interactions
in the Hamiltonian are short-ranged, and suggests that renormalization
techniques should be highly effective in this model.}
\end{figure}

To be in the universality class of the Ising model, we would expect
the non-local terms, $a_{l}$ to decrease exponentially with separation
$l$. Thus, when viewed at larger length scales, the non-local terms
would become irrelevant. The behaviour of $a_{n}$ for a variety of
$\theta=\chi$ is calculated numerically and presented in Fig. \ref{a_coeff}.
The case for $\theta\neq\chi$ is similar, and displays the same exponential
decrease in $a_{l}$ with $l$. Hence, we may naturally expect nearest-neighbour
interactions to be the most important interactions in this model -
an idea which we will make concrete in the following section.

Deriving an analytic expression for $a_{n}$ seems to be very difficult.
However it is possible to show a general dependence on $\theta$ and
$\chi$ of the form $\sin^{n}\theta\sin^{n}\chi$ for any particular
$n$. We can expand $\kappa_{k}$ in terms of $\eta_{k}$, where $\eta_{k}$
is defined in Eq. (\ref{eta_def}), as

\begin{equation}
\kappa_{k}=\sum_{p=0}^{\infty}\frac{2^{p-q}\Gamma^{2}(\frac{p+1}{2})}{\Gamma(p+1)}\eta_{k}^{p}=\sum_{p=0}^{\infty}c_{p}\eta_{k}^{p}\label{kappa_series}\end{equation}
where $\Gamma$ is the Euler gamma function. In turn, $\eta^{p}$
can be expressed as a series in terms of $\cos^{q}k$ as

\begin{eqnarray}
\eta_{k}^{p} & = & \sum_{q=0}^{p}\left(\begin{array}{c}
p\\
q\end{array}\right)(\cos\theta\cos\chi)^{p-q}(-\cos k\sin\theta\sin\chi)^{q}\label{eta_series}\\
 & = & \sum_{q=0}^{p}d_{p,q}\cos^{q}k\nonumber \end{eqnarray}

Finally, we can express $\cos^{q}k$ in terms of $\cos rk$. For the
case of $q$ even, this is

\begin{eqnarray}
\cos^{q}k & = & \sum_{r=0,r\in evens}^{q}\left(\begin{array}{c}
q\\
\frac{q-r}{2}\end{array}\right)\frac{\cos rk}{2^{q-1}}+c\label{cos_series}\\
 & = & \sum_{r=0}^{q}e_{q,r}\cos rk\nonumber \end{eqnarray}
where $c$ is an unenlightening constant, and a similar expression
holds for $p-q$ odd.

We can combine Equations (\ref{kappa_series}),(\ref{eta_series})
and (\ref{cos_series}) to yield an expression for $\kappa_{k}$

\begin{equation}
\kappa_{k}=\sum_{p=0}^{\infty}c_{p}\eta^{p}=\sum_{p=0}^{\infty}\sum_{q=0}^{p}\sum_{r=0}^{q}c_{p}d_{p,q}e_{q,r}\cos rk\label{eq:}\end{equation}
from which we can read the coefficient $a_{l}$ of $\cos lk$ as

\begin{equation}
a_{l}=\sum_{p=0}^{\infty}c_{p}\eta^{p}=\sum_{p=0}^{\infty}\sum_{q=0}^{p}c_{p}d_{p,q}e_{q,l}\label{eq:}\end{equation}

However, $e_{q,l}$ is only non-zero for $q>l$, and so we can replace
$\sum_{q=0}^{p}$ with $\sum_{q=l}^{p}$ to yield

\begin{eqnarray}
a_{l} & = & \sum_{p=0}^{\infty}c_{p}\eta^{p}=\sum_{p=0}^{\infty}c_{p}\sum_{q=l}^{p}d_{p,q}e_{q,l}\label{eq:}\end{eqnarray}
 which involves terms in $d_{p,q}\propto(\sin\theta\sin\chi)^{q}(\cos\theta\cos\chi)^{s}$
for $q>l$. Hence, for any given $a_{l}$, there is a behaviour proportional
to $(\sin\theta\sin\chi)^{l}$.

Of physical interest is the behaviour of $a_{l}$ with respect to
$l$ for a given $\theta$ and $\chi$. We will attempt to factor
out any behaviour in $l$ by noting that since $d_{pq},e_{q,l}<1$,
we can write

\begin{eqnarray}
a_{l} & = & \sum_{p=0}^{\infty}c_{p}\sum_{q=l}^{p}d_{p,q}e_{q,l}\label{eq:}\\
 & \leq & \sum_{p=0}^{\infty}c_{p}\sum_{q=l}^{p}d_{p,q}\sum_{q=l}^{p}e_{q,l}\nonumber \end{eqnarray}

However, if we turn our attention to the sum over $d_{p,q}$, we can
construct a further limit on $a_{l}$

\begin{eqnarray}
\sum_{q=l}^{p}d_{p,q} & = & \sum_{q=l}^{p}\left(\begin{array}{c}
p\\
q\end{array}\right)(\cos\theta\cos\chi)^{p-q}(-\cos k\sin\theta\sin\chi)^{q}\label{eq:}\\
 & \leq & \sum_{q'=0}^{p-l}\left(\begin{array}{c}
p\\
q'+l\end{array}\right)(\cos\theta\cos\chi)^{p-q'-l}(\sin\theta\sin\chi)^{q'+l}\label{eq:}\\
 & \leq & \sum_{q'=0}^{p-l}\left(\begin{array}{c}
p-l\\
q'\end{array}\right)(\cos\theta\cos\chi)^{p-l-q'}(\sin\theta\sin\chi)^{q'+l}\label{eq:}\\
 & \leq & (\sin\theta\sin\chi)^{l}(\cos\theta\cos\chi+\sin\theta\sin\chi)^{p-l}\label{eq:}\\
 & \leq & (\sin\theta\sin\chi)^{l}\label{eq:}\end{eqnarray}

Leading to our strictest inequality for $a_{l}$, showing an exponential
decay in $l$

\begin{eqnarray}
a_{l} & \leq & (\sin\theta\sin\chi)^{l}\sum_{p=0}^{\infty}c_{p}\sum_{q=l}^{p}e_{q,l}\label{eq:}\\
 & \leq & (\sin\theta\sin\chi)^{l}(\sum_{p=0}^{\infty}c_{p}p\max_{q}e_{q,l})\label{eq:}\end{eqnarray}

The case $\theta,\chi\rightarrow\frac{\pi}{2}$ has asymptotically
constant $a_{l}$. However, for $\theta,\chi\neq\frac{\pi}{2}$, we
can say that

\begin{eqnarray}
a_{l}\leq ke^{-\mu l}\rightarrow0 & \mathrm{\mathrm{as}} & l\rightarrow\infty\label{eq:}\end{eqnarray}
where $\mu=\ln(\sin\theta\sin\chi)$. Thus the terms in our model
$\bar{H}$ has only short range interactions.

\subsection{Renormalizing the Hamiltonian}

We have seen that the Hamiltonian, $\bar{H}$, associated with the
complete unitary $U$ is very complicated and involves non-local interactions.
Hence, we are forced to use renormalization group methods to extract
the interesting physics from this case.

Consider the continuum limit of the Hamiltonian, $\bar{H}$, which
is applicable in the thermodynamic limit. Near criticality, the physics
will be driven by long-wavelength effects, which suggest that the
wavevector $k$ will be small. The relevant excitations at low temperature
happen at the extremum of $\epsilon_{k}$, which we have shown occurs
at $k=0$. Hence, in the continuum limit, we can consider only low
lying states, near $k=0$. Under these approximations, our Hamiltonian
becomes

\begin{eqnarray}
\bar{H}\simeq\bar{H}' & = & \sum_{k}\bar{\kappa}\vec{\bar{\gamma}}_{k}(\chi,\theta).\vec{\nu}_{k}\label{eq:}\\
\vec{\bar{\gamma}}_{k} & = & (k\cos\theta\sin\chi,-k\sin\theta\sin\chi,\sin(\theta+\chi))\label{eq:}\\
\bar{\kappa} & = & \frac{\theta+\chi}{\sin(\theta+\chi)}\label{eq:}\end{eqnarray}
 which yields in terms if the fermion operators

\begin{eqnarray}
\bar{H}' & = & \sum_{k}ik\frac{(\theta+\chi)\sin\chi}{\sin(\theta+\chi)}(e^{i\theta}C_{k}^{\dagger}C_{-k}^{\dagger}+e^{-i\theta}C_{k}C_{-k})\label{eq:}\\
 &  & +2(\theta+\chi)C_{k}^{\dagger}C_{k}\nonumber \end{eqnarray}

Defining the continuum Fermi field\cite{Fields} as

\begin{equation}
\Psi(x_{i})=\frac{1}{\sqrt{a}}c_{i}\label{eq:}\end{equation}
where $a$ is the lattice spacing. $\Psi(x)$ satisfies the usual
anti-commutation relation $\{\Psi(x),\Psi^{\dagger}(x')\}=\delta(x-x')$.
Note that we can replace the sum over $k$ with an integral, by making
the substitution

\[
\sum_{k}a\rightarrow\int dx\]

Further, we expand the terms $C_{k}^{\dagger}C_{-k}^{\dagger}$ and
$C_{k}C_{-k}$ into terms of first order gradients $\frac{\partial\Psi^{\dagger}}{\partial x}$
and $\frac{\partial\Psi}{\partial x}$ through the use of identities
(Chapter 4 of Ref. \cite{Sachdev}). This yields

\begin{eqnarray}
\bar{H}'\simeq\tilde{H} & = & E_{0}+\int dx[\frac{(\theta+\chi)\sin\chi}{\sin(\theta+\chi)}(e^{i\theta}\Psi^{\dagger}\frac{\partial\Psi^{\dagger}}{\partial x}-e^{-i\theta}\Psi\frac{\partial\Psi}{\partial x})\label{eq:}\\
 &  & -2(\theta+\chi)\Psi^{\dagger}\Psi]\nonumber \end{eqnarray}
and $E_{0}$ is some constant.

Applying the transformation $\Psi\rightarrow e^{i\frac{\theta}{2}}\Psi$,
we have that

\begin{eqnarray}
\tilde{H} & = & E_{0}+\int dx[\frac{(\theta+\chi)\sin\chi}{\sin(\theta+\chi)}(\Psi^{\dagger}\frac{\partial\Psi^{\dagger}}{\partial x}-\Psi\frac{\partial\Psi}{\partial x})\label{eq:}\\
 &  & -2(\theta+\chi)\Psi^{\dagger}\Psi]\nonumber \end{eqnarray}

If we do not perform this transformation, we will have terms of the
form $\int dx\frac{\partial\Psi^{\dagger}}{\partial x}+\Psi\frac{\partial\Psi}{\partial x}$
in the Hamiltonian. One can show that these terms correspond to interactions
of the form $\sigma_{z}^{(n)}\sigma_{y}^{(n+1)}+\sigma_{y}^{(n)}\sigma_{z}^{(n+1)}$.
Chapter 4 of Ref. \cite{Fields} has further details regarding these
sorts of chiral symmetries in systems.

One can show that the Lagrangian corresponding to this Hamiltonian
will then be

\begin{eqnarray}
\mathcal{\tilde{L}} & = & \Psi^{\dagger}\frac{\partial\Psi}{\partial\tau}+\frac{(\theta+\chi)\sin\chi}{\sin(\theta+\chi)}(\Psi^{\dagger}\frac{\partial\Psi^{\dagger}}{\partial x}-\Psi\frac{\partial\Psi}{\partial x})\label{eq:}\\
 &  & -2(\theta+\chi)\Psi^{\dagger}\Psi\nonumber \end{eqnarray}
where $\tau$ is imaginary time.

Now we introduce the crucial step where we considering the effect
of scaling the problem. If we consider the effect of viewing the problem
at a scale $\delta^{l}$ more coarse in space, and $\delta^{zl}$
more coarse in time, we can introduce the new variables

\begin{eqnarray}
x' & = & x\delta^{-l}\label{eq:}\\
\tau' & = & \tau\delta^{-zl}\label{eq:}\\
\Psi' & = & \Psi\delta^{l/2}\label{eq:}\end{eqnarray}

We choose the value of the dynamic critical exponent, $z$, to be
identically equal to $1$, corresponding to an isotropy between space
and time, in order to leave the velocity-like coefficients of $\Psi^{\dagger}\frac{\partial\Psi^{\dagger}}{\partial x}$
and $\Psi\frac{\partial\Psi}{\partial x}$ unchanged.

For criticality to hold, these scaling conditions must leave the Lagrangian
unchanged. This occurs only when the quantity $\theta+\chi$ is identically
zero. This happens for $\theta=-\chi$, exactly as in the transverse
Ising model. 

Formally, if we write 

\begin{equation}
\tilde{\mathcal{L}}=\Psi^{\dagger}\frac{\partial\Psi}{\partial\tau}+u(\Psi^{\dagger}\frac{\partial\Psi^{\dagger}}{\partial x}-\Psi\frac{\partial\Psi}{\partial x})+\Delta\Psi^{\dagger}\Psi\label{eq:}\end{equation}
we will require that

\begin{eqnarray}
\Delta' & = & \Delta\delta^{l}\label{eq:}\\
u' & = & u\label{eq:}\end{eqnarray}
 implying that the scaling dimension of the term $\Delta\Psi^{\dagger}\Psi$,$\dim(\Delta)=1$.
Hence, $u$, and $\Delta$ are relevant parameters, having non-negative
scaling factors.

If we include second (or higher) order effects in $k$, we will include
terms of the form $\Delta'\Psi^{\dagger}\frac{\partial^{2}\Psi}{\partial x^{2}}$
or $\Delta''\Psi^{\dagger}\frac{\partial\Psi^{\dagger}}{\partial x}\frac{\partial\Psi}{\partial x}\Psi$
(or higher derivatives). From a simple analysis, one can show that
the parameters $\Delta'$ and $\Delta''$ are irrelevant, as the scaling
dimensions are\cite{Sachdev}

\begin{eqnarray*}
\dim(\Delta')=-1 & , & \dim(\Delta'')=-2\end{eqnarray*}

Recall the Ising chain in a transverse field, Eq. (\ref{ising}),
given by

\begin{eqnarray}
H_{Ising} & = & \sum_{n=1}^{N}\mu B\sigma_{x}^{(n)}+J\sigma_{z}^{(n)}\sigma_{z}^{(n+1)}\label{eq:}\\
 & = & \sum_{k}2BC_{k}^{\dagger}C_{k}+2J\cos kC_{k}^{\dagger}C_{k}\label{eq:}\\
 &  & +iJ\sin k(C_{k}^{\dagger}C_{-k}^{\dagger}+C_{k}C_{-k})\nonumber \end{eqnarray}
The Lagrangian for this model has the form

\begin{eqnarray}
\mathcal{L}_{Ising} & = & \Psi^{\dagger}\frac{\partial\Psi}{\partial\tau}+2(B+J)\Psi^{\dagger}\Psi-J(\Psi^{\dagger}\frac{\partial\Psi^{\dagger}}{\partial x}-\Psi\frac{\partial\Psi}{\partial x})\label{eq:}\end{eqnarray}

Hence, we may make an association between the two models through the
mapping

\begin{eqnarray}
J & = & -\frac{(\theta+\chi)\sin\chi}{\sin(\theta+\chi)}\label{eq:}\\
B+J & = & \theta+\chi\label{eq:}\end{eqnarray}

Thus we conclude that our continuum Hamiltonian $\tilde{H}$ belongs
in the same universality class as the transverse Ising Hamiltonian,
$H_{Ising}$. Hence it is possible to access the physical properties
at criticality of this well known model in a very straight forward
manner. The crucial assumptions which have been made are the assumption
of operation in the thermodynamic limit, and the low temperature (and
hence small $k$ excitation) regime, both of which are necessary for
renormalization to work. We will show in the next section that for
moderate $N$, the signatures of quantum phase transitions are still
observable.

\section{Signatures of a Quantum Phase Transition}

The first experimental realisations of a quantum simulations will
perhaps be seen on ion trap quantum computers. In this section we
review a few basic experimental signatures which may be seen in an
ion trap laboratory.

\subsection{Ground State Energy }

Recall that we have the following unitary map which describes the
system in terms of non-interacting fermions:

\begin{equation}
U(\chi,\theta)=e^{-i\sum_{k}\kappa_{k}\vec{\gamma}_{k}(\chi,\theta).\vec{\nu_{k}}}\label{eq:}\end{equation}

By inspection, the corresponding Hamiltonian is given by:

\begin{equation}
\bar{H}=\sum_{k}\kappa_{k}\vec{\gamma}_{k}(\chi,\theta).\vec{\nu_{k}}\label{eq:}\end{equation}

We note that eigenstates of $\bar{H}$ are simply products of the
eigenstates of $\bar{H}_{k}$, where $\bar{H}_{k}=\kappa_{k}\vec{\gamma}_{k}(\chi,\theta).\vec{\nu_{k}}$.
In the basis $\left|0\right\rangle ,C_{k}^{\dagger}\left|0\right\rangle ,C_{-k}^{\dagger}\left|0\right\rangle ,C_{k}^{\dagger}C_{-k}^{\dagger}\left|0\right\rangle $,
there will be four complex eigenvalues $\{\lambda_{k}^{(i)},i=1,2,3,4\}$,
with arguments $\{\omega_{k}^{(i)},i=1,2,3,4\}$. Let us define the
argument of a total system state, $\Omega^{(i)}=\sum_{k}\omega_{k}^{(i)}$,
where we form an eigenstate of $\bar{H}$ from matching eigenstates
of $\bar{H}_{k}$. Physically, we associate the argument of this state
with energy.

Now, we will consider the mapping

\begin{eqnarray*}
\chi\rightarrow r\cos\phi & , & \theta\rightarrow r\sin\phi\end{eqnarray*}
where $\phi$ can be considered as a relative strength between exchange
and field coupling terms, and $r$ is an overall strength. The Ising
criticality condition $\theta=\pm\chi$ is now $\phi=\pm\frac{\pi}{4},\pm\frac{3\pi}{4}$.

\begin{figure}
\includegraphics[%
  width=3.04414in]{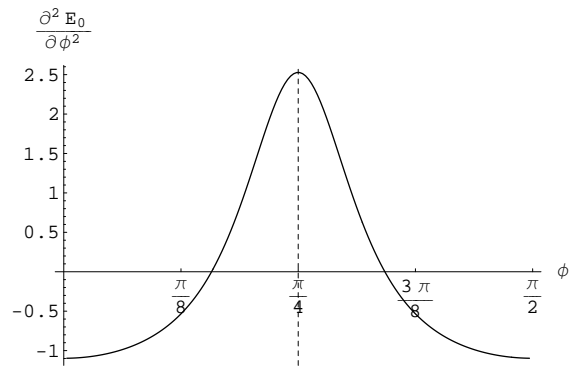}

\caption{\label{6q_d2phase}The second derivative of the ground state energy
of a 6 qubit model as a function of $\phi=\tan^{-1}\frac{\chi}{\theta}$,
for $r=1.9$. The maximum value is attained at $\phi=\frac{\pi}{4}$,
the value at which a quantum phase transition occurs in the thermodynamic
limit. While we see a strong maximum, in the thermodynamic limit,
we expect to see a singularity.}
\end{figure}

In Fig. \ref{6q_d2phase}, we observe an sharp peak in the second
derivative of the ground state energy, $\Omega^{(1)}$, with respect
to $\phi$. In the thermodynamic limit, this would become a singularity,
indicating a second order phase transition. Further, we observe that
this condition occurs for $\phi=\pm\frac{\pi}{4},\pm\frac{3\pi}{4}$,
which corresponds to the Ising transition.

We will demonstrate the nature of this singularity explicitly, by
noting that the argument of the ground state energy is

\begin{eqnarray}
\omega_{k}^{(1)} & = & -\kappa\sqrt{(\cos k\cos\theta\sin\chi+\sin\theta\cos\chi)^{2}+(\sin k\sin\chi)^{2}}\label{eq:}\\
 & = & \cos^{-1}\eta_{k}\equiv E_{k}\label{eq:}\end{eqnarray}

The next two eigenvalues are equal to $1$, and hence $\omega_{k}^{(2)}=\omega_{k}^{(3)}=0$.
The highest excited eigenstate has $\omega_{k}^{(4)}=-\omega_{k}^{(1)}$
by symmetry.

Substituting $\theta\rightarrow r\cos\phi$ and $\chi\rightarrow r\sin\phi$,
we evaluate

\begin{eqnarray}
\left.\frac{\partial^{2}\Omega^{(1)}}{\partial\phi^{2}}\right|{}_{\phi=\pm\frac{\pi}{4},\pm\frac{3\pi}{4}} & \simeq & \frac{2N}{\pi}\int_{0}^{\pi-\frac{\pi}{N}}\frac{(\cos k-1)dk}{\sqrt{1-(\cos^{2}\frac{r}{\sqrt{2}}-\cos k\sin^{2}\frac{r}{\sqrt{2}})^{2}}}\label{eq:}\end{eqnarray}

We find that the residue of the integrand is $\frac{4}{\sin r/\sqrt{2}}$,
and hence has a $\frac{1}{k}$ singularity. We can now conclude that
in the limit as $N\rightarrow\infty$, the value of $\left.\frac{\partial^{2}|\Omega^{(1)}|}{\partial\phi^{2}}\right|{}_{\phi=\pm\frac{\pi}{4},\pm\frac{3\pi}{4}}$
will be infinite, with a logarithmic singularity.

Alternatively, following Ref. \cite{bunder_mckenzie} and expressing
$E_{k}$ as\\
\begin{eqnarray}
E_{k} & = & \cos^{-1}\eta_{k}\label{omega_simple}\\
 & = & \cos^{-1}(\cos^{2}\frac{k}{2}\cos(\theta+\chi)+\sin^{2}\frac{k}{2}\cos(\theta-\chi))\nonumber \end{eqnarray}

Bunder and McKenzie\cite{bunder_mckenzie} note that for some wave
vector $k$, $E_{k}=0$, corresponding to a vanishing energy gap in
the system. When $E_{k}$ is expressed as Eq (\ref{omega_simple}),
it is clear that there will be no energy gap for $k=0$ if $\theta=-\chi$
and for $k=\pi$ if $\theta=\chi$. Without loss of generality, we
can consider the $k=0$ case, as the other is symmetric. Since the
relevant excitations are at $k\simeq0$, we may use Eq (\ref{omega_simple})
to expand $E_{k}^{2}$ as a series in $k$

\begin{eqnarray}
E_{k}^{2} & \simeq & (\theta+\chi)^{2}+k^{2}\frac{(\theta+\chi)(\cos\theta+\chi-\cos\theta-\chi)}{2\sin(\theta+\chi)}\label{eq:}\\
 & \equiv & \xi^{2}+k^{2}\zeta^{2}\nonumber \end{eqnarray}
and the ground state will have energy

\begin{eqnarray}
E_{0} & = & \frac{-1}{2\pi}\int_{-k_{c}}^{k_{c}}dk\sqrt{E_{k}^{2}}\label{eq:}\end{eqnarray}
where $k_{c}$ is a cutoff wavevector. While analytical solutions
are possible, it is unenlightening to solve this problem. Instead,
we can set out to determine the behaviour of the energy with respect
to a variable $\xi^{2}$ by using the indefinite integral

\begin{eqnarray}
E_{0} & = & -\int d\xi^{2}\frac{\partial}{\partial\xi^{2}}E_{0}\label{eq:}\end{eqnarray}

Carrying this out we obtain

\begin{eqnarray}
E_{0} & = & \int d\xi^{2}\frac{\partial}{\partial\xi^{2}}E_{0}\label{eq:}\\
 & = & \frac{-1}{4\pi}\int d\xi^{2}\int_{-k_{c}}^{k_{c}}dk\frac{1}{\sqrt{\xi^{2}+k^{2}\zeta^{2}}}\nonumber \\
 & = & \frac{-1}{4\pi}\int d\xi^{2}\frac{-2}{\zeta}(1-\ln\frac{\xi}{2\zeta k_{c}})\nonumber \\
 & = & \frac{-\xi^{2}}{2\pi\zeta}(1-2\ln\frac{\xi}{2\zeta k_{c}})\nonumber \end{eqnarray}

Thus we confirm the logarithmic nature of this singularity, and find
that $E_{0}\sim-\xi^{2-\alpha}$ where we have the value of the critical
exponent $\alpha=0^{+}$. This is the same behaviour as that found
in the transverse Ising model\cite{bunder_mckenzie}, which we expect
by their inclusion in the same universality class.

\begin{figure}
\includegraphics[%
  width=3.04414in]{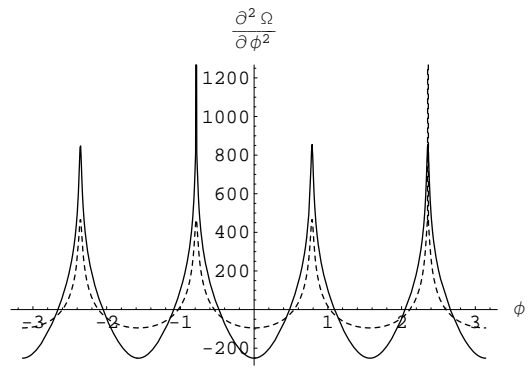}

\caption{\label{comparing}The second derivative of the phase of the eigenvalues
for the model Hamiltonian $\bar{H}$ (solid), based on Eq. (\ref{u_final}),
and for the transverse Ising Hamiltonian, $H_{Ising}$ (dashed), as
a function of $\phi=\tan^{-1}\frac{\chi}{\theta}$. Note that both
show singularities at $\theta=\pm\chi$. $\left|\theta^{2}+\chi^{2}\right|=r$
was chosen to be $1.9$ so as to highlight the differences between
the plots. For smaller $\theta$ and $\chi$, the commutator between
$H_{\theta}$ and $H_{\chi}$ becomes small, and the model becomes
asymptotically closer to the Ising model. Fig. \ref{a_coeff} shows
that the smaller $\theta$ and $\chi$, the faster the model's non-neighbour
parameters decay.}
\end{figure}

One can see the behaviour of ${\frac{\partial^{2}\Omega^{(1)}}{\partial\phi^{2}}}$
with respect to $\phi$ in Fig. \ref{comparing} for $N=200$ and
$r=1.9$. There exists a quantum phase transition at $\phi=\pm{\frac{\pi}{4}},\pm{\frac{3\pi}{4}}$
as evidenced by the singularity in ${\frac{\partial^{2}\Omega^{(1)}}{\partial\phi^{2}}}$.
This numerical modeling corresponds to our theoretical expectation
for the positions of the phase transitions. For a finite set of qubits,
one can clearly see the peak in the second derivative of the energy
with respect to $\phi$ in Fig. \ref{6q_d2phase}.

\subsection{Entanglement}

It has recently been shown that entanglement scales near a quantum
critical point \cite{Osborne,Osterloh}. Quantum phase transitions
are driven by quantum fluctuations\cite{Sachdev}, and entanglement
is a natural manner for non-local effects to manifest themselves.
As entanglement is a physical resource, it may be directly measured,
by a number of schemes\cite{measuring1,measuring2}. 

We denote the nearest neighbour entanglement in the ground state by
${\cal E}$. In the transverse Ising model, we see that the derivative
of entanglement with respect to $\phi$, $\frac{\partial{\cal E}}{\partial\phi}$,
near criticality to scale as a function of $\left|\phi-\phi_{c}\right|$.
Since we are in the same universality class, we expect to see identical
behaviour in this model. However, experimentally, isolating the ground
state of the system to observe this may be very difficult.

\begin{figure}
\includegraphics[%
  width=3.04414in]{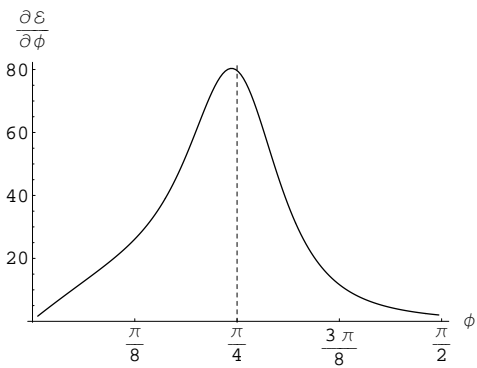}

\caption{\label{6q_entanglement}The derivative of the nearest neighbour entanglement,
with respect to $\phi$ for a 6 qubit model. Note that this has a
maximum very close to the critical point $\phi=\frac{\pi}{4}$. We
expect to see the nearest neighbour concurrence vary as $\log\left|\phi-\phi_{c}\right|$
in the thermodynamic limit.}
\end{figure}

\subsection{Spectroscopic Measurement of Eigenvalues}

Experimentally, it is possible that an ion trap may be used to implement
the unitary map of the form

\[
\left|\Psi\right\rangle \rightarrow e^{-i\chi\sum_{n=1}^{N}\sigma_{z}^{(n)}\sigma_{z}^{(n+1)}}\left|\Psi\right\rangle \equiv\left|\Psi'\right\rangle \]
followed by another map of the form

\[
\left|\Psi'\right\rangle \rightarrow e^{-i\theta\sum_{n=1}^{N}\sigma_{x}^{(n)}}\left|\Psi'\right\rangle \equiv\left|\Psi''\right\rangle \]

Let us introduce the notation

\[
\left|\Psi^{(m)}\right\rangle \equiv U^{m}\left|\Psi\right\rangle \]
where $\left|\Psi^{(m)}\right\rangle $ is the state after we repeat
this unitary map, $U$, $m$ times.

The state $\left|\Psi\right\rangle $ can be decomposed as

\begin{eqnarray}
U^{m}\left|\Psi\right\rangle  & = & \sum_{n}\ket{\phi_{n}}\bra{\phi_{n}}U^{m}\ket{\Psi}\label{eq:}\\
 & = & \sum_{n}\ket{\phi_{n}}\braket{\phi_{n}}{\Psi}e^{imE_{n}}\label{eq:}\end{eqnarray}
where $\left|\phi_{n}\right\rangle $ are the eigenstates of $U$,
with {}``energy'' $E_{n}$. Without loss of generality, we assume
these are ordered with

\[
E_{0}<E_{1}<\ldots<E_{M-1}\]
where $M=2^{N}$.

We can proceed to measure $U^{m}\left|\Psi\right\rangle $ in some
set of basis states, $\left|i\right\rangle $, which will typically
be binary computational basis states. Hence, we can measure

\begin{eqnarray}
\left|\bra{i}U^{m}\ket{\Psi}\right|^{2} & = & \left|\sum_{n}\braket{i}{\phi_{n}}\braket{\phi_{n}}{\Psi}e^{imE_{n}}\right|^{2}\label{eq:}\end{eqnarray}

\begin{equation}
=\sum_{n,n'}\braket{i}{\phi_{n}}\braket{i}\phi_{n'}^{*}\braket{\phi_{n}}{\Psi}\braket{\phi_{n'}}\Psi^{*}e^{im(E_{n}-E_{n'})}\label{eq:}\end{equation}

\begin{figure}
\includegraphics[%
  width=3.04414in]{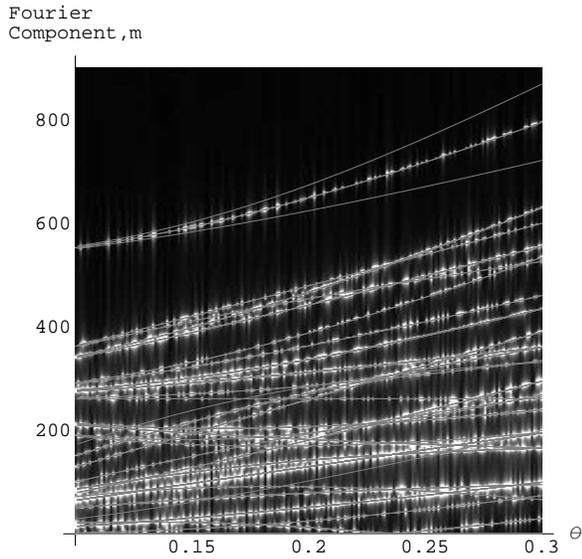}

\caption{\label{q4_fourier}The Fourier transform of $\left|\bra{i}U^{m}\ket{\Psi}\right|^{2}$,$F_{m,2048}$,
is shown as a vertical density, as function of the horizontal co-ordinate
$\theta$, for a fixed $\chi=0.2$, and 4 qubits. 4 qubits are chosen
so as to provide a complex, but not confusing diagram. The white bands
indicate a large Fourier component. The superimposed grey lines show
all energy differences - note that some of these are disallowed. In
this case, $2048$ samples are used in the Fourier series, and simulations
are taken in steps of $0.01$ in $\theta$. From this diagram, we
can see the energy gap between the ground state and first excited
state approaching zero. We can also see a level crossing, where one
of the grey lines is reflected through the origin at $\theta\simeq0.25$. }
\end{figure}

If we perform a Fourier transform of $\left|\bra{i}U^{m}\ket{\Psi}\right|^{2}$
over $m$, we expect to see peaks around the allowable transition
energies $E_{n}-E_{n'}$. A numerical simulation of this is shown
in Fig. \ref{q4_fourier} for 4 qubits. Let us define

\begin{equation}
F_{n}=\sum_{m=0}^{n-1}e^{\frac{i2\pi m}{n}}\left|\bra{i}U^{m}\ket{\Psi}\right|^{2}\label{eq:}\end{equation}
to be these Fourier components.

Since we are considering Unitary maps, and not Hamiltonians, we can
only determine the eigenvalues of $U$ to within an additive constant
of $2\pi$. Hence, in order that the Fourier components are not aliased
(that the energy levels do not {}``wrap around'' on themselves),
we require the ground state to have an energy $E_{0}>-\pi$, and the
highest excited state to have an energy $E_{2^{N}}<\pi$. Since the
energy is a function which scales with $O(N,\theta,\chi)$, we require
the condition $\max(\left|\theta\right|,\left|\chi\right|)<\frac{k_{int}}{N}$
where $k_{int}$ is $O(1)$. Keeping $\theta$ and $\chi$ small in
this manner will ensure that the energies will be resolvable uniquely
by the Fourier transform.

If we have a given set of energy eigenvalues, $\left\{ \bar{E}_{0},\bar{E}_{1}\ldots\bar{E}_{M-1}\right\} $,
we can form the set of energy differences, $\left\{ \bar{E}_{i,j}\equiv\bar{E}_{i}-\bar{E}_{j}\right\} $.
We can then calculate the Fourier transform of these differences,
and compare our measured spectrum with the calculated spectrum. If
we have $n\gg M$, then the problem is over determined, and we can
apply a least-squares method to reconstruct the original energy spectrum
(to within an additive constant, and global sign change). We may apply
the Levenberg Marquardt algorithm\cite{leven,MARQUARDT,LM} to perform
this reconstruction in polynomial time with an initial guess at the
set of energy eigenvalues.

\begin{figure}
\includegraphics[%
  width=3.04414in]{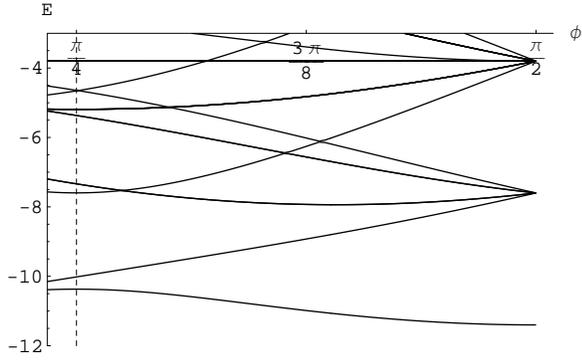}

\caption{\label{q6_eigens}Several of the lowest energy eigenvalues for the
6 qubit model. One can clearly see the excitation gap closing as $\phi$
approaches $\frac{\pi}{4}$. In the thermodynamic limit, we expect
the gap to be identically zero at $\phi=\frac{\pi}{4}$. However,
we can still observe the energy gap, $\Delta$, behaving as $\Delta\sim\left|\phi-\phi_{c}\right|^{\gamma}$
with some higher order corrections.}
\end{figure}

One could change the value of $\frac{\theta}{\chi}$ over many experiments
to tune the system through the critical coupling. In the thermodynamic
limit, the energy gap to the first excited state would vanish at criticality,
but we see in Fig. \ref{q6_eigens} that the condition is not strictly
met for a finite number of qubits.

One can use this to show that the energy gap, $\Delta$, for the excitation
from the ground to first excited state obeys the relation

\begin{equation}
\Delta\sim\left|\phi-\phi_{c}\right|^{\gamma}\label{eq:}\end{equation}
with $\gamma=1$, as we expect from a standard treatment of the transverse
Ising problem\cite{Sachdev}.

\subsubsection{Controlled-U Spectroscopy}

The above method requires knowledge of the approximate values of $\left|\bra{i}U^{m}\ket{\Psi}\right|^{2}$,
which means that a measurement with result $\ket{i}$ must be achieved
multiple times. In work by Miquel \emph{et al.}\cite{controlled_u_spec},
it has been shown that spectroscopy can be achieved much more easily
by implementing a controlled-U gate, and measuring only a single qubit\cite{one_bit}.
We can achieve this by using an ancillary qubit, and express the controlled-U
operation as

\[
CU:\bra{i}\otimes\bra{\Psi}\rightarrow\bra{i}\otimes U^{i}\bra{\Psi}\]
where $\bra{i}$ can be either $\bra{0}$, which takes $\bra{\Psi}$
to $\bra{\Psi}$, or $\bra{1}$, which takes $\bra{\Psi}$ to \textbf{$U\bra{\Psi}$.}

If we do a weak measurement on the control bit, we yield the result

\begin{eqnarray*}
\left\langle \sigma_{z}\right\rangle =\Re\left[Tr(U\rho)\right] & , & \left\langle \sigma_{y}\right\rangle =\Im\left[Tr(U\rho)\right]\end{eqnarray*}
where $\rho$ is the density matrix corresponding to the state has
been prepared in. If we prepare it in the mixed state given by $\rho=I/2^{n}$
where $I$ is the identity operator, we yield $\left\langle \sigma_{z}\right\rangle =\Re\left[Tr(U)\right]/N$,
which is proportional to the sum of the eigenvalues of $U$. If we
repeat this for $U^{m}$ for a variety of $m$, we can use the method
above to reconstruct the energy level diagram.

Further, Miquel \emph{et al.}\cite{controlled_u_spec} propose a scheme
using the quantum Fourier transform to probe specific regions of the
spectrum of the eigenvalues of $U$. This is achieved by introducing
an effective time scale into $U$, and exploiting the conjugacy of
energy and time.

\subsection{Phase Estimation Algorithm}

In work done by Abrams and Lloyd\cite{abramslloyd}, and further explored
in an ion trap context by Travaglione and Milburn\cite{eigenest},
it has been shown that it is possible to estimate the eigenvalues
associated with any unitary transformation, $U$. These correspond
directly to the energy eigenvalues of the equivalent Hamiltonian,
which we are interested in. The scheme also yields an approximate
eigenvector with high probability.

Starting with a mixed index state $\ket j_{I}$, and the state of
the target system $\ket{\Psi}$, we perform the transformation

\[
\Lambda(U):\ket j_{I}\ket\Psi_{T}\rightarrow\ket j_{I}\otimes U^{j}\ket\Psi_{T}\]
followed by a Fourier transformation on the index register. Measuring
the index register will then yield, with high probability, an approximate
eigenvector of $U$ in the target state, and information about the
phase of the eigenvalue of $U$ in the index register. 

Note, however, that use is made of an index register, which is at
least the same size as the system of interest. This makes it a much
more difficult problem to conquer experimentally, as a system twice
as big will be much more prone to decoherence. In ion trap implementations,
trapping twice as many ions will also be more difficult. While this
technique is superior to the spectroscopic measurements suggested
in the previous section, scalability issues may keep it from being
experimentally feasible for some time.

\section{Conclusion}

We have presented a number of key ideas which will drive our search
for a quantum phase transition in a system which is implementable
on an ion-trap quantum computer in a natural way.

We have taken the Feynman thesis and turned it around, to ask what
might happen if we have some implementable unitary transformation.
The two fields which will help to answer this question have been introduced
- namely, the ion-trap quantum architecture which will provide our
unitary transformation, and the tools of renormalization group theory.
In this paper, we have shown that the Hamiltonian corresponding to
a separated Ising map belongs to the same universality class as the
transverse Ising model. Further, the map presented here is realisable
in a very natural way on an ion-trap quantum computing architecture.

We have also suggested some experimental signatures , including ground
state energy and entanglement, and spectroscopic information which
may lead to the reconstruction of the energy spectrum.

After completion of this work, we became aware of some other work
on simulating quantum phase transitions in ion traps\cite{similar}

\begin{acknowledgments}
JPB would like to thank Ben Powell for useful discussions. This work
was supported by the Australian Research Council.
\end{acknowledgments}
\bibliographystyle{apsrev}
\bibliography{article1}

\end{document}